\documentstyle[psfig]{res}
\begin{document}

\title{Probing the width of the MACHO mass function}

\author{Anne M. GREEN \\
{\it Astronomy Unit, School of Mathematical Sciences, Queen Mary \\ 
and Westfield College, Mile End Road, London, E1 4NS, U.K.\\ 
amg@maths.qmw.ac.uk}}

\maketitle

\section*{Abstract}
The simplest interpretation of the microlensing events observed
towards the Large Magellanic Cloud is that approximately half of the
mass of the Milky Way halo is in the form of MAssive Compact Halo
Objects with $M \sim 0.5 M_{\odot}$.  This poses severe problems for
stellar MACHO candidates, and leads to the consideration of more
exotic objects such as primordial black holes (PBHs). Constraining the
MACHO mass function will shed light on their nature. Using the current
data we find, for four halo models, the best fit delta-function, power
law and PBH mass functions. The best fit PBH mass functions, despite
having significant finite width, have likelihoods which are similar
to, and for one particular halo model greater than, those of the best
fit delta functions. We also find that if the correct halo model is
known then $\sim$ 500 events will be sufficient to determine whether
the MACHO mass function has significant width, and will also allow
determination of the mass function parameters to $\sim 5\%$.

\section{Introduction}
The rotation curves of spiral galaxies are typically flat out to about
$\sim 30$ kpc. This implies that the mass enclosed increases linearly
with radius, with a halo of dark matter extending beyond the luminous
matter. The nature of the dark matter is unknown with possible
candidates including MAssive Compact Halo Objects (MACHOs) such as
brown dwarves, Jupiters or black holes and elementary particles, known
as Weakly Interacting Massive Particles (WIMPs), such as axions and
neutrilinos [2].

MACHOs with mass in the range $10^{-8} M_{\odot}$ to $10^{3}
M_{\odot}$ can be detected via the temporary amplification of
background stars which occurs, due to gravitational microlensing, when
a MACHO passes close to the line of sight to a background star
[9]. Since the early 1990s several collaborations have been monitoring
millions of stars in the Large and Small Magellanic Clouds, (LMC and
SMC), and a number of candidate microlensing events have been
observed.

The interpretation of these microlensing events is a matter of much
debate. Whilst the lenses responsible for these events may be located
in the halo of our galaxy, it is possible that the contribution to the
lensing rate due to other populations of objects has been
underestimated. For the standard--halo model, a cored isothermal
sphere, the most likely MACHO mass function (MF) is sharply peaked
around $0.5 M_{\odot}$, with about half of the total mass of the halo
in MACHOs. This poses a problem for stellar MACHO candidates (chemical
abundance arguments and direct searches place tight limits on their
abundance) [2], and leads to the consideration of more exotic MACHO
candidates such as primordial black holes (PBHs).

 PBHs with mass $M \sim 0.5 M_{\odot}$ could be formed due to a spike
in the primordial density perturbation spectrum at this scale or at
the QCD phase transition, where the reduced pressure forces allow PBHs
to form more easily [5]. In both cases it is not possible to produce
an arbitrarily narrow PBH MF and the predicted MF is considerably
wider than the sharply peaked MFs which have been fitted to the
observed events to date [1].

\section{Current data}
In their analysis of their 2-year data the MACHO collaboration form a
6 event sub-sample which they argue is a conservative estimate of the
events resulting from lenses located in the Milky Way halo [1]. We
find the maximum likelihood fit, to the 6 event `halo sub--sample',
for DF, power law and PBH MFs for four sample halo models: the
standard--halo, the standard--halo including the transverse velocity
of the line of sight and 2 power--law halo models [3].  For the
standard--halo, both with and without the transverse velocity of the
line of sight, and one of the power--law halo models, the DF MF has
the largest maximum likelihood, whilst for the other power--law halo
the PBH MF provides the best fit.

The differences in maximum likelihood between MF/halo model
combinations are small and, unsurprisingly given the small number of
events, it is not possible to differentiate between MFs using the
current data, even if the halo model is fixed (for more details see
[4]).

\section{Monte Carlo simulations}
We carried out 400 Monte Carlo simulations each for $N=100,316,1000$
and 3162 event samples, assuming a standard--halo and taking the MACHO
MF to be the, comparatively, `broad' PBH MF. For each simulation we
found the best fit PBH and delta-function (DF) MFs.

For each simulation we compared the theoretical event rate
distributions produced by the best fit PBH and DF MFs with those
`observed' using a modified form of the Kolmogorov-Smirnov (KS)
test. The fraction of the simulations passing the KS test at a given
confidence level, for both the PBH and DF MACHO MFs, is shown in
Fig. 1 for each $N$.

In Fig. 2 we plot 1 and 2 $\sigma$ contours (which contain 68\% and
95\% of the simulations respectively), of the mean MACHO mass and halo
fraction, of the best fit PBH MFs. The values for the input MF are
marked with a cross. We find that fitting a DF MF when
the true MF is the PBH MF leads to a systematic underestimation of the
mean MACHO mass by $\sim 15 \%$. For further details see ref. [4].
  
\begin{figure}[t]
\centerline{\psfig{figure=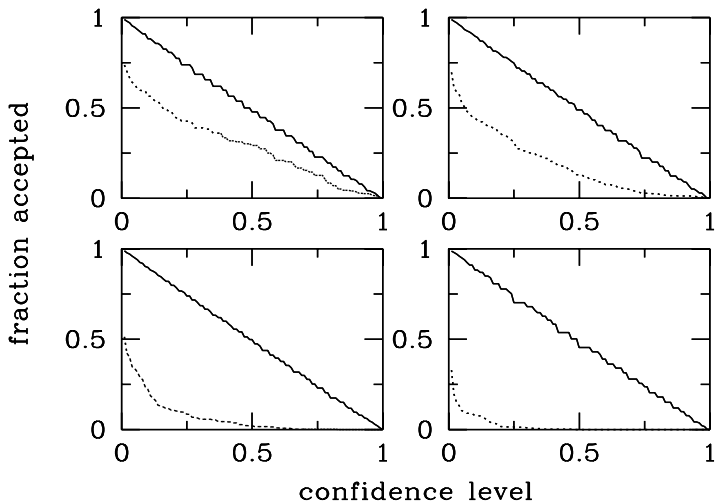,width=10cm,height=7cm,angle=0}}
\caption{The fraction of best fit PBH (solid lines) and DF
(dotted) MFs passing the KS test at a given confidence
level, for $N=100, 316, 1000$ and 3162 events from right to left, 
upper row then lower row.}
\label{fig.1}
\end{figure}

\begin{figure}[t]
\centerline{\psfig{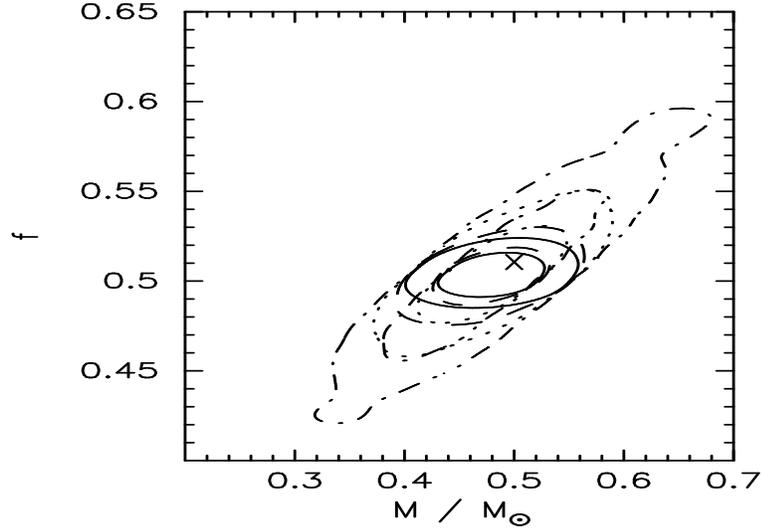}}
\caption{Contours containing 68 per-cent and 95 per-cent, of the halo
fraction and mean MACHO mass, of the best fit PBH MFs, for $N= 3162$
(solid line), $1000$ (dashed), $316$ (dotted) and $100$ (dot-dashed)
events.}
\label{fig.2}
\end{figure}

\section{Future prospects}

Assuming that the lenses are located in the Milky Way halo, and that
the correct halo model is known, approximately 500 events should be
sufficient to ascertain whether the MACHO mass function has
significant finite width, and also determine the parameters of the mass
function to $\sim 5\%$. If the halo model is not known then the number
of events necessary is likely to be increased by a least an order of
magnitude [6], however the use of a satellite to make parallax
measurements of microlensing events would allow simultaneous
determination of the lens location and, if appropriate, the halo
structure and mass function with of order 100s events [7]. If the
MACHOs are PBHs then the gravitational waves emitted by PBH-PBH
binaries will allow the MACHO mass distribution to be mapped by the
Laser Interferometer Space Antenna [8].

\vspace{1pc}

\section*{References}

\re
1.  Alcock\, C.\, et. al. 1997, ApJ, 486, 697
\re
2. see for instance \ Carr\, B. J. in these proceedings
\re
3. Evans\, N.\, W.\, 1993, MNRAS, 260, 191
\re
4. Green\, A.\, M.\, 1999 preprint astro-ph/9912424
\re
5. Jedamzik\, K.,\, \& Niemeyer\, J.\,C.\, 1999 Phys. Rev. D 
59, 124014
\re
6. Markovic\, D.,\, \& Sommer-Larsen\, J.\, 1997, MNRAS, 229, 929
\re
7.\ Markovic\, D.\, 1998 MNRAS, 507 316
\re
8.\ Nakamura\, T.\, Sasaki\, M.\, Tanaka\, T.,\, \& Thorne\, K.\, 
                  1997 ApJ 487 L139 and Ioka\, K. in these proceedings
\re
9.\ \ Paczy\'{n}ski\, B. 1986, ApJ, 428 L5

%

\end{document}